\documentclass[a4paper]{article}
\usepackage{WCSB,amsmath,epsfig}

\title{MICROTUBULE TRACKING FROM STOCHASTIC OPTICAL RECONSTRUCTION MICROSCOPY IMAGES}
\name{Juliane Liepe$^{1}$, Federico Felizzi$^{2}$, Agata Pernus$^{3}$,  Maria Hanulova$^{4}$}

\address{$ ^1$  \normalsize{Biochemistry Building, Department of Biological Sciences, Imperial College London, SW7 2AZ London, UK}\\
$ ^2$\normalsize{D-BSSE, ETH Zurich, Mattenstrasse 26, 4058 Basel, Switzerland}, \\
$ ^3$\normalsize{DKFZ, Division Biophysics of Macromolecules, Im Neuenheimer Feld 580, D-69120 Heidelberg, Germany}\\ 
$ ^4$\normalsize{Universit\"at Bayreuth, Lehrstuhl Experimentalphysik I, Universit\"atsstrasse 30, D-95447 Bayreuth, Germany}\\
juliane.liepe08@imperial.ac.uk, federico.felizzi@bsse.ethz.ch\\a.pernus@dkfz.de,  	maria.hanulova@uni-bayreuth.de}

\begin{document}

\maketitle
\begin{abstract}
Our work aims at using quantitative imaging tools to complement the
limitation of noise encountered by high resolution fluorescence microscopy
methods. Several cycles of fluorophore activation, imaging and deactivation produce a
sequence of images in which the signals of individual fluorophores do not
overlap, due to the low light intensity during their activation. The centroid position of each fluorophore is then determined by Gaussian fitting of each signal, where the final resolution depends on the precision with which each fluorphore is localized.
Superimposing the images will result in having the same fluorophore mapped
onto a `cloud' of locations. The most significant information of the
superimposed images is contained in the macro-structures identifying
microtubules, mitochondria or other organelles. Cascades of binary image
processing algorithms are applied in order to isolate the larger organelles.
A Markovian algorithm selecting the nearest neighbour is finally applied to
the de-noised images, to automatically extract relevant information on microtubules.
Our work supplements advancements in experimental technologies with computational methods, helping quantifying sub-cellular properties with high accuracy.

\end{abstract}
\section{INTRODUCTION}
\label{sec:intro}
Fluorescence microscopy is an important tool in understanding the structure of living cells. It presents an intrinsic limitation due to the diffraction of light. Fluorophores which are too close will be mapped to identical images, making it impossible to resolve details at nanoscopic scales. Stochastic optical reconstruction microscopy (STORM) \cite{rust2006} provides a technology for high precision localisation of single molecules. At every imaging cycle, only a small percentage of molecules is switched on, so that their images do not overlap. 
STORM imaging has been applied to the identification of microtubules \cite{Bates:2007p15442}, rendering visible clear-cut filamentous structures. Major attention of previous work \cite{Bates:2009p15424} was given to the isolation of clusters of points away from the filamentous structures, whose analysis is significant to the resolution resulting from the imaging technology. Microtubules are polymeric subcellular components of the cytoskeleton. Their dynamics is characterized by the stochastic transition between extension and shrinking phases \cite{Karsenti:2006p15508}. Understanding the molecular properties of microtubule polymers is a crucial step towards the elucidation of the causes of microtubule dynamic instability and - at an increasing level of complexity - of the entire cellular achitecture \cite{Karsenti:2006p15508}. The notion of persistence length provides a measure of the binding energy of one-dimensional polymers, resulting in different responses to stochastic thermal fluctuations \cite{Gutjahr:2006p15539}. The isolation of filamentous structures from STORM imaging allows for the application of tracking algorithms to elucidate on the physical properties of cytoskeletal components. In \cite{Koulgi:2010p15581} a HMM was proposed to track the curvilinear structures resulting from imaging of microtubules. 
In this work we propose an approach that combines nearest neighbor information and directionality to track multiple filamentous structures. Our algorithm exhibits a tremendous performance in the tracking of intersecting microtubules. 
\cite{Gonzalez:1998p13605} provides a valuable reference for digital image processing algorithms. \cite{Zhuang:2009p14308}\cite{Koulgi:2010} \cite{Karsenti:2006p15508}

\section{EXPERIMENTAL SETTING}
\subsection{Laser}
\subsection{Microscope}
\subsection{Software}

\section{METHODS}
\label{sec:general}

\subsection{Imaging}
The raw STORM image results from 500-100000 image frames. Initially, each individual image frame is convolved with a Gaussian filter. The filter, with the same width as the point spread function of the microscope, acts as a band-pass filter, as it removes high frequency noise and low-frequency variations of the background. Threshdolding the image results in the identification of local maxima, which are annotated as peaks. 

\subsection{Image processing}
\label{ssec:ImageProc}
The raw data representing the detected thresholded peaks are the coordinate points of about 240000 locations, presenting a location precision of $10^{-3}$ pixels, which would result in a 256000x256000 pixel image. For our microtubule tracking purposes, we introduce a hard-cut on the resolution, so that there is a 1-1 mapping between the raw locations and the final binary image. After testing various image sizes, we found that a 4000x4000 pixel image turns out to achieve a good matching between the raw locations and pixels.  
\subsubsection{Isolating microtubules}
\label{sssec:IsolMic}
\begin{figure}[!ht]
\centering
\includegraphics[width=7cm, height=6.5cm]{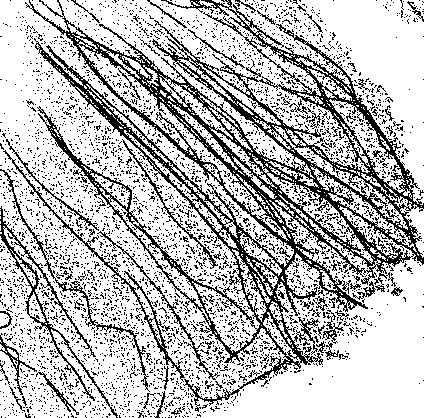}
\caption{Raw 4000x4000 pixel STORM image. The image shows a section of a cell with labelled microtubules.}
\label{Noised_Image}
\end{figure}

We use the described methodology to identify microtubule structures. The image resulting from the preliminary processing described in \ref{ssec:ImageProc} results in the presence of a number of peaks that do not belong to any microtubule. In order to train an algorithm that identifies and tracks the microtubule correctly, we devised a method that removes the noise from the image (figure \ref{Noised_Image}). 
Our method consists of a series of morphological operations, outlined in the following list
\begin{itemize}
\item perform a dilation $ OR \left\lbrace W(f(x,y)) \right\rbrace $, where the window $W(f(x,y))$ is defined by a round-shaped structuring element of radius $3$.
\item perform a small region removal, that discards all the clusters containing less than 500 pixels. 
\item apply a \textit{majority filtering} twice . A 3x3 window was used for the majority filtering operation. 
\end{itemize}
\begin{figure}[!ht]
\centering
\includegraphics[width=8cm, height=6.5cm]{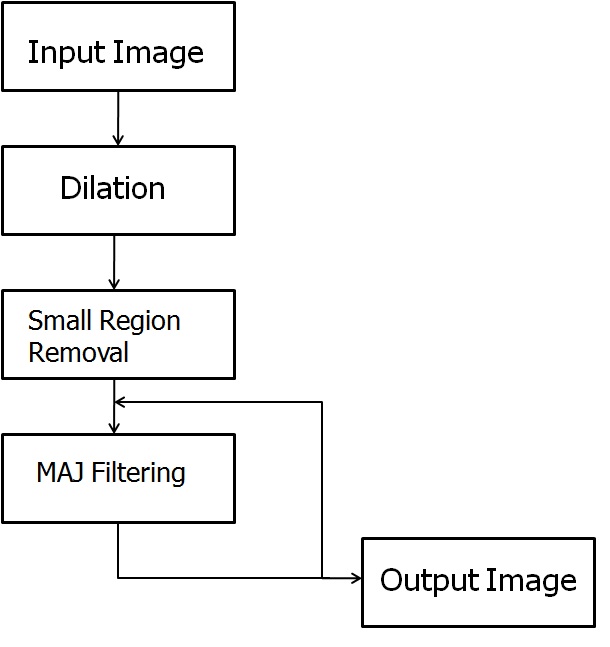}
\caption{Block structure of the image processing algorithm to isolate the microtubules from background noise.}
\label{fig:Flow}
\end{figure}
\begin{figure}
\centering
\includegraphics[width=8cm, height=6.5cm]{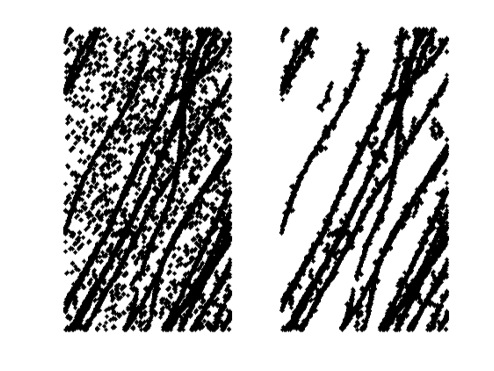}
\caption{Noise reduction. Shown is a section of the raw image. Additionally to the microtubules (connected black lines) microtubule-unrelated signals are present in the image (left). Application of our algorithm (section \ref{sssec:IsolMic}) results in removing of the noise (right).}
\label{fig1}
\end{figure}

\subsection{Microtubule Tracking Algorithm}
Let $\mathcal{S}$ be the set of all the points in the denoised picture. The algorithm starts by selecting one point $s_0 \in \mathcal{S} $ at random. Let $\mathcal{N}_{s_0}$ be the neighborhood of $s_0$. $\mathcal{N}_{s_0}$ is defined as 
\begin{equation}
\mathcal{N}_s := \{x : ||x - s||_{2} \leq d_{\mbox{max}} \}
\end{equation}
where $d_{\mbox{max}}$ is varied at each iteration, so that the cardinality $|\mathcal{N}_s |$ is kept fixed.  
The point $s_0$ is initially marked as visited. The selection of the subsequent points involves the construction of a Markov chain $\{ X_i\}$. The states of the Markov chains are locations of the points on the image. The transition probabilities are defined according to two criteria. One involves the proximity of the next point to be selected to the point defining the current state of the chain. The other criterion aims at keeping the smoothness of the curve resulting from the union of the visited points. The first criterion reads: 
\begin{equation}
\label{eqn:markov1}
P \left(X_i = s_i| X_{i-1} = s_{i-1} \right) \propto d ( s_i, s_{i-1} )^{-1}
\end{equation}
For the second criterion, we define $\mathcal{V}$ to be the set of all visited locations. If the cardinality of $\mathcal{V}$ is larger than a threshold value, we define the smooth line $\mathcal{L}$ interpolating the points in $\mathcal{V}$. The successive state in the Markov chain is selected by assigning a penalty score on the points in $\mathcal{N}_{X_i}$ which are away from the line $\mathcal{L}$:
\begin{equation}
\label{eqn:markov2}
P \left(X_i = s_i| X_{i-1} = s_{i-1} \right) \propto d ( s_i, \mathcal{L} )^{-2}
\end{equation}
The introduction of the criterion \ref{eqn:markov2} is fundamental when the current state of the Markov chain happens to be on the intersection of two or more microtubules. The application of condition \ref{eqn:markov1} only would not discriminate the privileged direction that the microtubule being traced follows. 
\begin{figure}
\centering
\includegraphics[width=5cm, height=3.5cm]{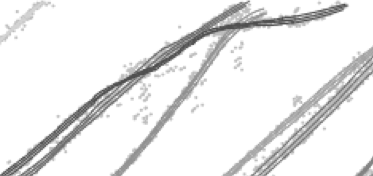}
\caption{Detail on the intersection of distinct microtubules. Our approach keeps track of the line interpolating the points tracked and discriminate the right bend of the microtubule being traced. Different levels of gray represent different traced microtubules}
\label{fig1}
\end{figure}

\section{RESULTS}

\subsection{Isolation of microtubules}

Using our tracking algorithm we were able to isolate microtubules and their characteristics (figure \ref{fig1}).
In our approach we assume that microtubules have a long persistence length, i.e. they are mainly straight in the cell. With this assumption we are able to track overlapping microtubules. However, because we only favour straight microtubules, but do not exclude the possibility of curved shapes, the algorithm still manages to track non-straight microtubules.

\subsection{Persistence Length}
\label{ssec:perlen}
\begin{figure}
\centering
\includegraphics[width=7cm, height=5.5cm]{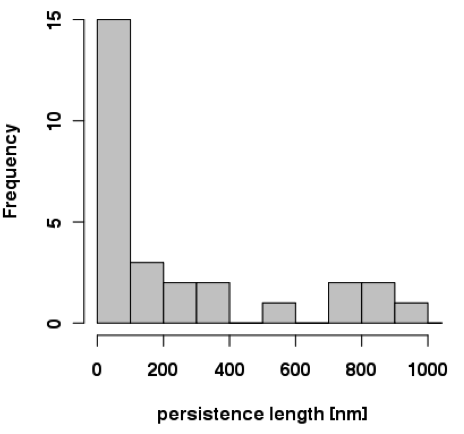}
\caption{Histogram for the detected persistence lengths}
\label{fig:perlen}
\end{figure}

An important characteristics of semi-flexible polymers is persistence length $l_p$. Persistence length is a measure on the stiffness of a polymer. It measures the correlation of the directions of the tangent vectors along the line defining the polymer. Let $r_0, r_1, \ldots, r_N$ be points on a curve tracking a microtubule. Let us define the increments $\{ \Delta r_i \} = \left( r_1 - r_0, \ldots, r_N - r_{N-1}\right)$. The spatial correlations
\begin{equation}
\langle \Delta r_i, \Delta r_j \rangle
\end{equation}
decay exponentially as 
\begin{equation}
\exp \lbrace \frac{-|i - j|} {l_p} \rbrace
\end{equation}
We computed persistence lengths for the tracked microtubules, and identified an average persistence length of $292.8nm$. The distribution of persistence lengths is shown in figure \ref{fig:perlen}. 

\section{CONCLUSION}
Starting from technological advancements in high resolution microscopy, we processed and analyzed the resulting images. 
The combination of our noise reduction algorithm with the tracking algorithm allowed us to automatically isolate microtubules and to extract their characteristics. These algorithms will prove very helpful for quantitative image analysis as manual microtubule tracking is too time consuming and error prone depending on the experience of the user. Our algorithm can distinguish between overlapping microtubules in a 2D plane. However, as for most image processing algorithms, the quality of the raw image will strongly influence the error rate of the tracking results. Too noisy raw data can lead to incomplete isolated microtubules.

These algorithms can be applied to gain more detailed information about microtubules. Here, we analysed the persistence length as an application example. Further characteristics as the width and its variation along each microtubule can be computed from the extracted data. More complex analysis would include the packaging of the microtubules in an entire cell and the microtubule density.

With time resolved STORM images our algorithms could be applied to investigate microtubule dynamics. A future challenge is the implementation of these algorithms to allow for online image reconstruction and analysis.

\section{ACKNOWLEDGMENTS}
We would like to thank Melike Lakadamyali and Graham Dempsey for their precious support while we set up the experiments and designed the image processing and microtubule tracking algorithms. 

\bibliographystyle{IEEEbib}
\bibliography{Microtubule_Tracking}

\begin{thebibliography}{1}

\bibitem{rust2006}
M.~J. Rust, M.~Bates, and X.~Zhuang,
\newblock ``Stochastic optical reconstruction microscopy (storm) provides
  sub-diffraction-limit image resolution,''
\newblock {\em Nat Methods}, vol. 3, pp. 793--795, 2006.

\bibitem{Bates:2007p15442}
M.~Bates and et. al.,
\newblock ``Multicolor super-resolution imaging with photo-switchable
  fluorescent probes,''
\newblock {\em Science}, vol. 317, pp. 1749--1753, 2007.

\bibitem{Bates:2009p15424}
M.~Bates and et. al.,
\newblock ``Sub-diffraction-limit imaging with stochastic optical
  reconstruction microscopy,''
\newblock {\em Nobel Volume on Single Molecule Spectroscopy in Chemistry},
  2009.

\bibitem{Karsenti:2006p15508}
E.~Karsenti, F.~Nedelec, and T.~Surrey,
\newblock ``Modelling microtubule patterns,''
\newblock {\em Nature Cell Biology}, vol. 8, pp. 1204--1211, 2006.

\bibitem{Gutjahr:2006p15539}
P.~Gutjahr, R.~Lipowsky, and J.~Kierfeld,
\newblock ``Persistence length of semiflexible polymers and bending rigidity
  renormalization,''
\newblock {\em Europhysics Letters}, 2006.

\bibitem{Koulgi:2010p15581}
P.~Koulgi, M.~E. Sargin, K.~Rose, and B.~S. Manjunath,
\newblock ``Graphical model-based tracking of curvilinear structures in
  bio-image sequences,''
\newblock in {\em Proceedings of the 2010 20th International Conference on
  Pattern Recognition}, Washington, DC, USA, 2010, ICPR '10, pp. 2596--2599,
  IEEE Computer Society.

\bibitem{Gonzalez:1998p13605}
R.~C. Gonzales and R.~E. Woods,
\newblock {\em Digital Image Processing},
\newblock Prentice Hall, 2001.

\bibitem{Zhuang:2009p14308}
X.~Zhuang,
\newblock ``Nano-imaging with storm,''
\newblock {\em Nature photonics}, vol. 3, pp. 365 -- 367, 2009.

\end{thebibliography}

\end{document}